\documentclass[12pt]{article}
\pdfoutput=1
\usepackage{amsmath,amssymb,amsfonts,epsfig,graphicx,euscript}%
\usepackage[pdftex,bookmarks,bookmarksnumbered,linktocpage,pdfstartview=FitH]{hyperref}
\hypersetup{colorlinks,%
citecolor=red,%
filecolor=blue,%
linkcolor=blue,%
urlcolor=blue,%
pdftex}
\usepackage[all]{hypcap}
\numberwithin{equation}{section}
\usepackage{cite}
\makeatother

\newcommand{\bse}{\begin{subequations}}
\newcommand{\ese}{\end{subequations}}
\newcommand{\be}{\begin{equation}}
\newcommand{\ee}{\end{equation}}
\newcommand{\bea}{\begin{eqnarray}}
\newcommand{\eea}{\end{eqnarray}}
\newcommand{\ba}{\begin{array}}
\newcommand{\ea}{\end{array}}

\begin{document}
\hfill%

\begin{center}
{\Large{\bf Algebraic Form of M3-Brane Action}}

\vskip .5cm {\large Hossein Ghadjari} \vskip .1cm {\it Department of
Physics, Amirkabir University of Technology
(Tehran Polytechnic)\\
P.O.Box 15875-4413, Tehran, Iran}\\
{\sl h-ghajari@aut.ac.ir}\\
\vskip .5cm {\large Zahra Rezaei} \vskip .1cm {\it Department of
Physics, University of Tafresh
\\
P.O.Box 39518-79611, Tehran, Iran}\\
{\sl z.rezaei@aut.ac.ir}\\
\end{center}

\vspace{0cm}
%\bigskip
\begin{abstract}
%\textbf{Abstract}
We reformulate the bosonic action of unstable M3-brane to manifest
its algebraic representation. It is seen that in contrast with
string and M2-brane actions that are represented only in terms of
two and three dimensional Lie-algebras respectively, the algebraic
form of M3-brane action is a combination of four, three and two
dimensional Lie-algebras. Corresponding brackets appear as mixtures
of tachyon field, space-time coordinates, $X$, two-form field,
$\hat{\omega}^{(2)}$, and Born-Infeld one-form, $\hat{b}_\mu$.

\end{abstract}



\newpage

\tableofcontents

\section{Introduction}

Algebraic reformulation of known actions in string theory and
M-theory shows that string theory is based on conventional algebra
or two dimensional Lie-algebra (known as two-algebra) but a complete
description of M-theory needs an extended Lie algebra called
three-algebra \cite{Lee1} which was mainly developed by Bagger,
Lambert and Gustavsson \cite{BL1, BL2, BL3, BLG}. Numbers two and
three are associated with string theory and M-theory, respectively.
Two is the string worldsheet dimension and also the codimension of
D-branes in both type IIA and IIB superstring theories
\cite{polchinski}. Three is the membrane worldvolume dimension in
M-theory and the codimension of M2 and M5-branes. It means that via
two-algebra interactions some Dp-branes will condense to a
D(p+2)-brane \cite{Myers} and through three-algebra interactions
multiple M2-branes condense to a M5-brane
\cite{Matsuo,Imamura,Krishnan,Jeon1,Park,Bandos1,Bandos2,Jeon2,Lee2}.
These connections between two and three and respectively string
theory and M-theory become obvious by rewriting Nambu-Goto actions
in algebraic form.

By analogy one can expect to describe p-branes applying p+1-algebra
structure \cite{Kamani}. These extended algebras are applied to
construct worldvolume theories for multiple p-branes in terms of
Nambu brackets that are classical approximations to multiple
commutators of these algebras \cite{Nambu}. Nambu n-brackets
introduce a way to understand $n$ dimensional Lie-algebra presented
by Fillipov \cite{Fillipov}. Formulation of p-brane action in terms
of p+1-algebra makes it more compact and we are left with algebraic
calculations that are usually simpler to handle.

Since in string theory we are, inevitably, faced with unstable
systems, study of them deepens our understanding of string theory.
In bosonic string theory the instability is always present due to
tachyon presence in open string spectrum. Two examples of unstable
states in superstring theories are: non-BPS branes (odd (even)
dimensional branes in type IIA (IIB) theory) and brane-anti-brane
pairs in both type IIA and IIB theories \cite{Sen1,Sen2}. One of the
interesting facts about the dynamics of these unstable branes,
generally obvious in effective action formulation, is their
dimensional reduction through tachyon condensation
\cite{Kutasov,Lee3,Hashimoto,Lerda,rezaei1,rezaei2}. During this
process the negative energy density of the tachyon potential at its
minimum point, cancels the tension of the D-brane (or D-branes)
\cite{Sen3}, and the final product is a closed string vacuum without
a D-brane or stable lower dimensional D-branes. On the other hand
stable objects in string theory can be obtained by dimensional
reduction of stable branes in M-theory (M2 and M5-branes).
Naturally, one can expect to have a pre-image of unstable branes in
superstring theories by formulating an effective action for unstable
branes in M-theory. Among different unstable systems in M-theory
\cite{IK} M3-brane is noteworthy because it is directly related to
M2-brane. Tachyon condensation of the M3-brane effective action
results in M2-brane action and also its dimensional reduction leads
to non-BPS D3-brane action in type IIA string theory \cite{Kluson}.

Despite attempts made to formulate M3-brane action consistent with
desired conditions \cite{Kluson} there has been no algebraic
approach towards this formulation. Existence of algebraic form for
the action of M2-brane, as the fundamental object of M-theory,
motivated us to search for the algebraic presentation of M3-brane as
the main unstable object in M-theory that its instability is due to
the presence of tachyon.

What distinguishes present study from conventional algebraic
formulations is instability of M3-brane. In other words, presence of
tachyon and other background fields affect the resultant algebra. It
is shown that pure four-algebra does not occur, as expected, and we
are encountered with four, three and two-brackets that are mixtures
of tachyon, spacetime coordinates and other fields.

\section{Algebraic M3-brane action}

The conventional action corresponding to a non-BPS M3-brane is a
combination of DBI (Dirac-Born-Infeld) and WZ (Wess-Zumino) parts
\cite{Kluson}
\be %
\label{eq1}
\begin{split} %
S&= {S_{DBI}} + {S_{WZ}}, \cr%
{S_{DBI}}&=-{\int{{d^4}\xi V(T){|\hat{k}|}^{1/2}}\sqrt {-\det H_{\mu\nu}}}, \cr%
{S_{WZ}}&=-\int{{d^4}\xi} V(T)\varepsilon
^{\mu_1\mu_2\mu_3\mu_4}{\partial_{\mu_1}}T
\hat{\kappa}_{\mu_2\mu_3\mu_4}, \cr%
\end{split}\ee %
where $\xi^\mu$ with $\mu=0,1,2,3$ label worldvolume coordinates of
M3-brane. $V(T)$ is the tachyon potential which is an even function
of $T$ and is characterized as $V(T=\pm\infty)=0$ and
$V(T=0)={{\cal{T}}_{M3}}$ where ${{\cal{T}}_{M3}}$ is M3-brane
tension. $\hat{k}^M (X)$ is the Killing vector and the Lie
derivative of all target space fields vanish with respect to it
\cite{Kluson}. Other fields in (\ref{eq1}) are defined as
\be \label{eq2}%
\begin{split} %
{H_{\mu\nu }}&={\hat{g}}_{MN}{\hat{D}}_\mu \hat{X}^M \hat{D}_\nu
\hat{X}^N +\frac{1}{|\hat{k}|}\hat{F}_{\mu\nu}+
\frac{1}{|\hat{k}|}\partial_\mu T \partial_\nu T, \cr%
 \hat{k}^2&= \hat{k}^M \hat{k}^N \hat{g}_{MN},\;\;\;\; \hat{k}^2 =|\hat{k}|^2, \cr%
 \hat{F}_{\mu\nu}&=\partial_\mu \hat{b}_\nu-\partial_\nu
\hat{b}_\mu
+\partial_\mu \hat{X}^M \partial_\nu \hat{X}^N (i_{\hat{k}}\hat{C})_{MN}, \cr%
 \hat{D}_\mu X^M &= \partial_\mu \hat{X}^M-\hat{A}_\mu
\hat{k}^M,\;\;\;\;\hat{A}_{\mu}=\frac{1}{{|\hat{k}|}^2}\partial_\mu \hat{X}^M \hat{k}_M, \cr%
 \hat{\kappa}_{\mu_2\mu_3\mu_4}&=
\partial_{\mu_2}\hat{\omega}_{\mu_3\mu_4}^{(2)}-\partial_{\mu_3}\hat{\omega} _{\mu_2\mu_4}^{(2)}+\partial_{\mu _4}\hat{\omega}_{\mu_2\mu_3}^{(2)} \cr%
 &+ \frac{1}{3!}\hat{C}_{KMN}\hat{D}_{\mu_2}\hat{X}^K
\hat{D}_{\mu_3}\hat{X}^M\hat{D}_{\mu_4}\hat{X}^N+\frac{1}{2!}\hat{A}_{\mu_2}(\partial_{\mu_3}\hat{b}_{\mu_4}-\partial_{\mu_4}\hat{b}_{\mu_3}).\cr%
\end{split}
\ee %
The tensor $H_{\mu\nu}$ consists of the pullback of background
metric, field strength $\hat{F}_{\mu\nu}$ of gauge field $A_\mu$ and
tachyon field, $T$. $M$ and $N$ represent spacetime indices and
$\hat{D}_{\mu}$ is covariant derivative. The field strength itself
is expressed in terms of Born-Infeld 1-form $\hat{b}_\mu$ and R-R
sector field $\hat{C}$. The curvature of the 2-form
$\hat{\omega}^{(2)}$ is shown as $\hat{\kappa}$.

Determinant of the tensor $H_{\mu\nu}$ in DBI action can be
decomposed as
\be\label{det} %
\sqrt{-\det H_{\mu\nu}}=\sqrt{-\det
(\tilde{G}_{\mu\nu}+\tilde{F}_{\mu\nu})},
\ee %
where
\bea %
&~&\tilde{F}_{\mu\nu}=\partial_\mu \hat{b}_\nu -\partial_\nu \hat{b}_\mu, \nonumber\\
&~&\tilde{G}_{\mu\nu}=L_{MN}\partial_\mu X^M \partial_\nu
X^N+\frac{1}{|\hat{k}|}\partial_\mu T \partial_\nu T,
\eea %
and
\be %
L_{MN}=g_{MN}+\frac{(i_{|\hat{k}|}\hat{C}_{MN})}{|\hat{k}|}-\frac{\hat{k}_M
\hat{k}_N}{|\hat{k}|^2}.
\ee %
Regarding (\ref{det}), DBI action can be expanded to quadratic order
\cite{becker} as
\be\label{dbiexpansion} %
S_{DBI}=-\int d^4 \xi V(T) \sqrt{-\det
\tilde{G}_{\mu\nu}}\left(1+\frac{1}{4}\tilde{F}_{\mu\nu}\tilde{F}^{\mu\nu}+...
\right).
\ee %

\subsection{DBI part of M3-brane action}

To find the algebraic form of the DBI action, we start with the
first term in (\ref{dbiexpansion}), i.e.
$\sqrt{-\det{\tilde{G}_{\mu\nu}}}$, that is determinant of a
$4\times4$ matrix and all its elements are sum of a tachyonic part
and a space-like part $(\partial X \partial X+\partial T \partial
T)$. This determinant is totally consisted of $48\times 8$ terms.
These terms can be classified into sixteen $4\times4$ determinants
in such a way that the elements of these determinants are only
$\partial X
\partial X$ or $\partial T
\partial T$ and not sum of them. So each determinant has $24$ terms
that summing them up leads to the same number of terms
($16\times24$) as the initial main determinant. These $16$
determinants can be categorized as: one determinant with $\partial X
\partial X$ elements (four combinations from 4
states $ \left(
\begin{array}{rcl}
4 \\
4
\end{array}\right)=1.
$), one determinant with elements of $\partial T \partial T$ ($
\left(
\begin{array}{rcl}
4 \\
4
\end{array}\right)=1$), four determinants with three rows of $\partial X \partial
X$ elements and one row of $\partial T \partial T$ elements ($
\left(
\begin{array}{rcl}
4 \\
1
\end{array}\right)=4$), four determinants with three rows of $\partial T \partial
T$ elements and one row of $\partial X \partial X$ elements ($
\left(
\begin{array}{rcl}
4 \\
1
\end{array}\right)=4$) and finally six determinants with two rows of
$\partial T \partial T$ elements and two rows of $\partial X
\partial X$ elements ($ \left(
\begin{array}{rcl}
4 \\
2
\end{array}\right)=6$). It is obtained that
determinants with more than one row of $\partial T \partial T$ are
zero. So we are left with two kinds of determinants: a determinant
consisting of only $\partial X
\partial X$ entities and those with three rows of $\partial X
\partial X$ elements and one row of $\partial T \partial T$
entities. Since determinant does not change under exchanging of
rows, by considering all possible permutations ($4!$) of rows for
each one of the remaining determinants, the form of the
four-algebra, in accordance with (\ref{nambu}), emerges. At the end
of the day after a tedious calculation the algebraic form of
$\sqrt{-\det{\tilde{G}_{\mu\nu}}}$ is obtained as
\bea\label{algebraic-det} %
\sqrt{-\det{\tilde{G}_{\mu\nu}}}&\rightarrow&\bigg{\{}-
\bigg{(}L_{MN}L_{OP}L_{QR}L_{ST}[X^M,X^O,X^Q,X^S][X^N,X^P,X^R,X^T]\cr %
&+&\frac{4}{|\hat{k}|}L_{MN}L_{OP}L_{QR}[T,X^M,X^O,X^Q][T,X^N,X^P,X^R]
\bigg{)} \bigg{\}}^{1/2}.
\eea %
The 4-bracket of spacetime coordinates, $X$'s, corresponds to
algebraic action derived in \cite{Kamani,Lee1} for $p=3$ case and
with the fermionic fields turned off. The new term here is the mixed
four-bracket of $X$'s and $T$.

Presenting a general algebraic form for the term
$\tilde{F}_{\mu\nu}\tilde{F}^{\mu\nu}$ in DBI action is not
possible, however in some special cases it finds a simple form. For
example one can consider a selfdual (anti-selfdual) field strength
that corresponds to instanton. An instanton is a static (solitonic)
solution to pure Yang-Mills theories \cite{VV}. They are important
in both supersymmetric field theories and superstring theories
mostly because of their nonperturbative effects. They also play role
in M-theory for instance in applying the M2-brane actions to
M5-brane \cite{chu}. The solution to field equations in Yang-Mills
theory corresponding to an instanton has a selfdual (anti-selfdual)
field strength \cite{VV}. Considering this property gives the
following expression for
$\rm{tr}\;\tilde{F}_{\mu\nu}\tilde{F}^{\mu\nu}$ in the case of
regular one-instanton solution \cite{VV}
\be \label{trace}%
\rm{tr}\;\tilde{F}_{\mu\nu}\tilde{F}^{\mu\nu}=-96\frac{\rho^4}{((x-x_0)^2+\rho^2)^4},
\ee %
where $x_0$ and $\rho$ are arbitrary parameters called collective
coordinates.

So for the instantonic case the full algebraic form of the DBI part
of the action reads as
\bea\label{algebraic-dbi} %
S_{DBI}&=&-\int d^4 \xi
V(T)\left(1-24\frac{\rho^4}{((x-x_0)^2+\rho^2)^4}
\right)\cr%
&\times&\bigg{\{}-
\bigg{(}L_{MN}L_{OP}L_{QR}L_{ST}[X^M,X^O,X^Q,X^S][X^N,X^P,X^R,X^T]\cr %
&+&\frac{4}{|\hat{k}|}L_{MN}L_{OP}L_{QR}[T,X^M,X^O,X^Q][T,X^N,X^P,X^R]
\bigg{)} \bigg{\}}^{1/2}.
\eea %

\subsection{WZ part of M3-brane action}

The integrand of WZ action in (\ref{eq1}) can be divided into three
parts by replacing $\hat{\kappa}$ from (\ref{eq2})
\bea\label{ws} %
S_{WZ}&\rightarrow&\varepsilon^{\mu_1\mu_2\mu_3\mu_4}\partial_{\mu_1}T\hat{\kappa}_{\mu_2\mu_3\mu_4}\cr %
&=&\varepsilon^{\mu_1\mu_2\mu_3\mu_4}\partial_{\mu_1}T\bigg{(}
\partial_{\mu_2}\hat{\omega}_{\mu_3\mu_4}^{(2)}-\partial_{\mu_3}\hat{\omega} _{\mu_2\mu_4}^{(2)}
+\partial_{\mu _4}\hat{\omega}_{\mu_2\mu_3}^{(2)}\cr %
 &+& \frac{1}{3!}\hat{C}_{KMN}\hat{D}_{\mu_2}\hat{X}^K
\hat{D}_{\mu_3}\hat{X}^M\hat{D}_{\mu_4}\hat{X}^N\cr%
&+&\frac{1}{2!}\hat{A}_{\mu_2}(\partial_{\mu_3}\hat{b}_{\mu_4}-\partial_{\mu_4}\hat{b}_{\mu_3})\bigg{)},
\eea %
and each part is dealt with separately.

By expanding the first part, three terms of $\hat{\omega}^{(2)}$
derivatives, and considering all possible permutations of
four-dimensional Levi-Civita symbol,
$\varepsilon^{\mu_1\mu_2\mu_3\mu_4}$, we come to a view of
two-algebra. The reason is that according to (\ref{nambu}) having
two derivative factors signals a two-algebra which carries its
two-dimensional Levi-Civita symbol. But since here only different
permutations of $\varepsilon^{\mu_1\mu_2\mu_3\mu_4}$ give correct
signs to the terms, multiplying the resultant two-algebra by another
two-dimensional Levi-Civita symbol and using the relation
\[
\varepsilon^{\alpha\beta}\varepsilon_{\gamma\delta}=\delta^{\alpha}_{\gamma}
\delta^{\beta}_{\delta}-\delta^{\alpha}_{\delta}\delta^{\beta}_{\gamma},
\]
conduct us to the correct form. So the first part of WZ action is
reformulated in terms of two-bracket as
\bea\label{firsttermws} %
S_{WZ,1}&\rightarrow&\varepsilon^{\mu_1\mu_2\mu_3\mu_4}\partial_{\mu_1}T(
\partial_{\mu_2}\hat{\omega}_{\mu_3\mu_4}^{(2)}-\partial_{\mu_3}\hat{\omega} _{\mu_2\mu_4}^{(2)}
+\partial_{\mu _4}\hat{\omega}_{\mu_2\mu_3}^{(2)})\cr %
&=&3
\varepsilon^{\mu_1\mu_2\mu_3\mu_4}\varepsilon_{\mu_1\mu_2}[T,\omega_{\mu_3\mu_4}].
\eea %

In the second part of WZ action, three $X$ derivatives, $\partial
X$, and one tachyon derivative, $\partial T$, appear in a way that
obviously form a four-algebra
\bea\label{secondtermws} %
S_{WZ,2}&\rightarrow&\frac{1}{3!}\hat{C}_{KMN}\varepsilon^{\mu_1\mu_2\mu_3\mu_4}\partial_{\mu_1}T\hat{D}_{\mu_2}\hat{X}^K
\hat{D}_{\mu_3}\hat{X}^M\hat{D}_{\mu_4}\hat{X}^N\cr %
&=&\frac{1}{3!}\hat{C}_{KMN}\varepsilon^{\mu_1\mu_2\mu_3\mu_4}\left(1-\frac{\hat{k}^P\hat{k}_P}{|\hat{k}|^2}\right)^3
\partial_{\mu_1}T\partial_{\mu_2}X^K\partial_{\mu_3}X^M\partial_{\mu_4}X^N\cr%
&=&\frac{1}{3!}\hat{C}_{KMN}\left(1-\frac{\hat{k}^P\hat{k}_P}{|\hat{k}|^2}\right)^3[T,X^K,X^M,X^N].
\eea %

Substituting $A_{\mu}$ in the last part of WZ action we are faced
with terms consisting of $\partial X$, $\partial T$ and $\partial b$
that according to (\ref{nambu}) indicate a three-algebra. Similar to
the argument made for the first part of WZ action, multiplying this
three-bracket by a three-dimensional Levi-Civita symbol and using
the identity
\[
\varepsilon^{\alpha\beta\gamma}\varepsilon_{\delta\eta\lambda}=\delta^\alpha_\delta
(\delta^\beta_\eta \delta^\gamma_\lambda-\delta^\beta_\lambda
\delta^\gamma_\eta)-\delta^\alpha_\eta(\delta^\beta_\delta
\delta^\gamma_\lambda-\delta^\beta_\lambda
\delta^\gamma_\delta)+\delta^\alpha_\lambda(\delta^\beta_\delta
\delta^\gamma_\eta-\delta^\beta_\eta \delta^\gamma_\delta),
\]
give the convenient three-algebra. Different permutations of
four-dimensional Levi-Civita symbol are responsible for correct
signs of different terms in three-algebra. So the algebraic form of
this part is
\bea\label{tirdtermws} %
S_{WZ,3}&\rightarrow&\frac{1}{2!}\varepsilon^{\mu_1\mu_2\mu_3\mu_4}\partial_{\mu_1}T
\hat{A}_{\mu_2}(\partial_{\mu_3}\hat{b}_{\mu_4}-\partial_{\mu_4}\hat{b}_{\mu_3})\cr %
&=&\frac{\hat{k}_M}{2!|\hat{k}|^2}\varepsilon^{\mu_1\mu_2\mu_3\mu_4}\varepsilon_{\mu_1\mu_2\mu_3}[T,X^M,b_{\mu_4}].
\eea %

Therefore WZ action of M3-brane is presented in terms of two, three
and four-brackets as
\bea\label{algebraicws} %
S_{WZ}&=&-\int d^4 \xi
V(T)\bigg{\{}3 \varepsilon^{\mu_1\mu_2\mu_3\mu_4}\varepsilon_{\mu_1\mu_2} [T,\omega_{\mu_3\mu_4}]\cr %
&+&\frac{1}{3!}C_{KMN}\left(1-\frac{\hat{k}^P\hat{k}_P}{|\hat{k}|^2}\right)^3[T,X^K,X^M,X^N]\cr %
&+&\frac{\hat{k}_M}{2!|\hat{k}|^2}\varepsilon^{\mu_1\mu_2\mu_3\mu_4}\varepsilon_{\mu_1\mu_2\mu_3}[T,X^M,b_{\mu_4}]\bigg{\}}.
\eea %
It is seen that tachyon field couples with spacetime coordinates,
Born-Infeld one-form $\hat{b}_\mu$ and two-form $\hat{\omega}^{(2)}$
through four, three and two-brackets, respectively.

\section{Summary and conclusion}

In this article we presented an algebraic form for bosonic M3-brane
action by reformulating this action in terms of brackets. Since in
the literature p-branes are described by p+1-algebra \cite{Kamani}
one expects a four-algebra structure for M3-brane. But it was shown
that the algebraic representation of M3-brane is a combination of
four, three and two-algebras. Generally this difference stems from
the instability of the system that tachyon is responsible for.
Except of a four-bracket of spacetime coordinates in DBI part,
tachyon field is present in all other brackets and forms four, three
and two-brackets with spacetime coordinates, two-form,
$\hat{\omega}^{(2)}$, and Born-Infeld one-form $\hat{b}_\mu$,
respectively . In future we try to study the dimensional reduction
of this algebraic action.

\appendix
\section{Fillipov n-Lie algebra} \label{App:AppendixA}

Fillipov n-Lie algebra \cite{Fillipov} as a natural generalization
of a Lie algebra is defined by n-bracket satisfying the totally
antisymmetric property
\be\label{antisym} %
[X_1,...,X_i,...,X_j,...,X_n]=-[X_1,...,X_j,...,X_i,...,X_n],
\ee%
and the Leibniz rule
\be\label{Leibniz} %
[X_1,...,X_{n-1},[Y_1,...,Y_n]]=\sum_{j=1}^{n}[Y_1,...,[X_1,...,X_{n-1},Y_j],...,Y_n].
\ee%
n-Lie algebra is equipped with an invariant inner product
\be\label{innerproduct} %
\langle X,Y \rangle=\langle Y,X \rangle,
\ee%
as well as the invariance under the n-bracket transformation
\be\label{trans} %
\langle [X_1,...,X_{n-1},Y],Z \rangle+\langle
Y,[X_1,...,X_{n-1},Z]\rangle=0.
\ee%
When $n=2$ the definition reduces to the usual Lie algebra and the
inner product can be given by "Trace".

n-Lie algebra can be realized in terms of Nambu n-bracket defined
over functional space on an n-dimensional manifold \cite{Nambu}
\be\label{nambu} %
[X_1,X_2,...,X_n]\Leftrightarrow\{X_1,X_2,...,X_n\}_{N.B}:=\frac{1}{\sqrt{{\cal{G}}}}
\epsilon^{l_1l_2...l_n}\partial_{l_1}X_1\partial_{l_2}X_2...\partial_{l_n}X_n.
\ee %
${\cal{G}}$ is determinant of the metric of the manifold and can be
chosen arbitrarily since the properties
(\ref{antisym})-(\ref{trans}) hold irrespective of the presence of
the local factor \cite{Lee1}.

\end{document}